\documentclass[12pt]{iopart}

\usepackage[english]{babel}
\usepackage{amsfonts}
\usepackage{graphicx}
\usepackage{color}

\newcommand{\ket}[1]{|{#1}\rangle}

\newcommand{\braket}[2]{\langle{#1}|{#2}\rangle}
\newcommand{\ave}[1]{\langle{#1}\rangle}
\newcommand{\qave}[3]{\langle{#1}|{#2}|{#3}\rangle}

\newcommand{\andid}[1]{{\color{black}#1}}

\newcommand{\an}[1]{{\color{black}#1}}
\newcommand{\refa}[1]{{\color{black}#1}}

\begin{document}

\title[]{\andid{Errors in quantum optimal control and strategy for the search of easily implementable control pulses}}

\author{Antonio Negretti$^1$, Rosario Fazio$^{2,3}$, and Tommaso Calarco$^1$}

\address{
1. Institute for Quantum Information Processing, University of Ulm, Albert-Einstein-Allee 11, D-89069 Ulm,
Germany.\\
2. NEST, Scuola Normale Superiore and Istituto di Nanoscienze - CNR,
Pisa, Italy.\\
3. Center for Quantum Technologies, National University of Singapore, Republic of Singapore
}

\begin{abstract}
We introduce a new approach to assess the error of control problems we aim to optimize. 
The method offers a strategy to define new control \an{pulses} that are not necessarily optimal but 
still able to yield an error not larger than some fixed a priori threshold, and therefore provide control 
pulses that might be more amenable for an experimental implementation. The formalism is applied 
to an exactly solvable model and to the Landau-Zener model, whose optimal control problem is 
solvable only numerically.  The presented method is of importance for applications where 
a high degree of controllability of the dynamics of closed quantum systems is required. 
\end{abstract}

\maketitle


\section{Introduction}

The theory of optimal control (OC) that has been mathematically formulated
in the last century by the seminal works of Pontryagin, Bellman, 
Kalman, Stratonovich \cite{Krotov1996,Doherty2000,Maybeck1994} has been 
instrumental for the achievement of highly reliable electronic devices 
used to control, for instance, mechanical systems such as airplanes, cars, 
etc., but also to control chemical reactions or to design ultra-fast laser 
pulses for manipulating molecules (e.g., to break a certain bond while 
leaving other bonds intact \cite{Brixner2004}), and today even to 
optimize (stochastic) financial analyses \cite{Oksendal2000}. 

The topic has recently attracted the attention of physicists working in 
quantum information and computation science, because of the need to 
engineer accurate protocols. To this aim different numerical techniques 
have been devised in order to minimize (or maximize) some performance 
criterion or, alternatively called, objective functional. We mention the most used 
ones in open-loop quantum control: the 
Krotov iterative method \cite{Krotov1996,Sklarz2002,Reich2010} and the gradient ascent 
pulse engineering algorithm \cite{Khaneja2005}. Although these are powerful 
tools for the search of OC pulses they do not provide an assessment of \andid{the 
tolerable error against distortions and do not guarantee to obtain control pulses 
easily realizable in the laboratory.} Such an issue is of paramount importance 
for quantum information processing (QIP), since the error allowed by 
fault-tolerant quantum computation ranges 
between 0.01\% to fractions of a percent \cite{Steane2003,Knill2005}. 

A possible (empirical) approach relies on applying an arbitrary distortion to 
the OC solution and then looking at the error that it produces on the objective 
functional, or by selecting a region in the Hilbert space that is robust against 
noise (decoherence free-subspace) \cite{Zanardi1997}, whose existence 
follows from the symmetry properties of the noise. Recently, a more systematic 
methodology based on the Hessian analysis of the cost functional has been 
proposed \cite{Ho2009} \an{or by using an improved genetic algorithm in the 
presence of control noise \cite{Shuang2004}.}

\andid{The aim of this work is to provide an alternative method to evaluate to 
which extent a given OC scheme can tolerate errors.} The method is based on 
the Hessian approximation of the cost functional, as in Ref. \cite{Ho2009}, but it is 
applicable to any \andid{system} Hamiltonian $\hat{H}(u_t)$ and Hilbert space dimension. 
Such arbitrariness is important in several circumstances where either the control 
pulse $u_t\equiv u(t)$ does not appear 
linearly in $\hat{H}(u_t)$ or where the state to be controlled is an auxiliary state (e.g., 
the motional state of an atom \cite{Treutlein2006}) and not the quantum bit itself (e.g., 
an atomic internal state). The control of such states is relevant not only for several QIP 
implementations, but also for quantum metrological purposes, where the control of 
large quantum superpositions may increase the sensitivity of precise measurements. 

Even though usually it is not possible to know
analytically the OC pulse, we underscore that our
\andid{assumption} is that the parameter obtained with some numerical
algorithm is very close to the global optimum. More precisely, 
the error on the cost functional obtained with the numerically
found OC pulse has to be much smaller than the error allowed by the process we
are interested to optimize. Besides the interest on its own, 
we believe that our approach \andid{might be} of importance for experiments, where, 
typically, optimal pulses are extremely difficult to achieve. 
To this aim, our method could help to find easily implementable 
control signals \an{(EICS)}, while still being able to satisfactorily 
fulfill the performance criterion we are interested in.
\an{Here with ``easily implementable control signals" reference is made to pulses that can be 
utilized to control an experiment at the quantum level. More precisely, since nowadays the experiments 
are typically controlled by a computer, an obvious requirement for the control signal is that its Fourier 
spectrum has to match the bandwidth of the transducer or it can not vary faster 
than the clock frequency of the processor. Besides this, since the computer during the 
course of the experiment controls some device (e.g., the applied voltage on 
electrodes or electric current \cite{Treutlein2006,Boehi2009}) the control pulse has, for instance, to take into account the bandwidth of those devices. 
These conditions might be not satisfied by the optimal control pulse obtained with the aforementioned 
optimization algorithms. Even though, recently, some extensions of those optimization methods 
in order to include spectral constraints on the control pulses have been made \cite{Gollub2008,Lapert2009}, these are not 
always easy to be handled, especially when the dynamics of a many-body quantum system is concerned.}

\section{The method}

\begin{figure}[t]
\begin{center}
\includegraphics[width=6cm,height=3cm]{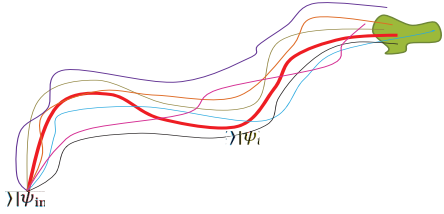}
\end{center}
\vspace{-0.7cm}
\caption{(Color online) Pictorial representation of the optimal
  trajectory $\ket{\psi_t^{\mathsf{o}}}$ (red-thick line) obtained with the OC
  pulse $u_t^{\mathsf{o}}$ and trajectories (thin lines) for
  non-optimized control pulses. The shaded
  (green) area represents the portion of Hilbert space within which
  the cost functional $\mathsf{J}\le\mathcal{J}$ (see also text), that
  is, the subset of state vectors close to the goal state $\ket{\psi_{\mathrm{g}}}$.}
\label{fig:sketch}
\end{figure}
Let us briefly review what is the purpose and what are the methodologies
adopted commonly in the quantum control research area concerning closed
quantum systems and the link between OC theory and analytical mechanics. The usual 
problem is the engineering of a \andid{system} Hamiltonian $\hat{H}(u_t)$ such that 
the initial state $\ket{\psi_{\mathrm{in}}}$ at time $t=t_0$ of the quantum
system under consideration is brought at time $t=T$ to some desired goal
state $\ket{\psi_{\mathrm{g}}}$ (see also Fig. \ref{fig:sketch}). To this aim there 
are two (equivalent) techniques for searching optimal pulses at our 
disposal: the variational method and dynamical programming. In both cases 
the goal is the minimization of the cost functional 

\begin{eqnarray}
\mathsf{J}[t_0,u_t,\psi_t] = \mathsf{G}(\psi_T) 
+ \int_{t_0}^T\mathrm{d}t \mathsf{C}(t,u_t,\psi_t)
\label{eq:Jdef}
\end{eqnarray}
over all admissible control pulses $u_t$ and state trajectories 
$\ket{\psi_t} \equiv \ket{\psi(t)}$ given a certain initial
state $\ket{\psi_{\mathrm{in}}}$. 
Here by admissible we mean any $u_t$ for which the Schr\"odinger equation is
well-defined and has a unique solution $\ket{\psi_t}$ given the initial
condition $\ket{\psi_{\mathrm{in}}}$. The first term on the r.h.s. of 
Eq.~(\ref{eq:Jdef}) is the terminal functional, e.g., the overlap infidelity $1 -
\vert\braket{\psi_{\mathrm{g}}}{\psi_T}\vert^2$. The second term provides 
 additional constraints on the control pulse~\footnote{For instance, 
the ``laser electric field fluence'' with $\mathsf{C}(u_t) = u^2_t$, where $u_t$ is an electric field 
amplitude.}. In the variational 
 method the additional constraint $\mathrm{Re}\{\int_{t_0}^T
\mathrm{d}t[\braket{\chi_t}{\dot{\psi}_t}+i\qave{\chi_t}{\hat{H}(u_t)}{\psi_t}]\}$
is introduced \cite{Peirce1988}, where we set $\hbar\equiv 1$ and $\ket{\chi_t}$ is a Lagrange multiplier often
referred to as costate, which ensures that the state $\ket{\psi_t}$ satisfies the 
Schr\"odinger equation. The search of an extremum of the cost
functional produces a set of equations for the state, costate and the 
control pulse. 

On the other hand, dynamical programming, based on the Bellman's
optimality principle \cite{Kirk2004}, produces an equation, the so called 
Hamilton-Jacobi-Bellman equation, for the optimal cost 
$\mathsf{S}(t,\psi_t):=\inf_{u_t}\mathsf{J}[t,u_t,\psi_t]$. This equation is
expressed in formally the same way as the Hamilton-Jacobi equation of 
classical mechanics, but with the difference that it is propagated backwards 
in time. The role of the Hamilton function in classical mechanics is played 
in control theory by the (quantum) Pontryagin
Hamiltonian $\mathsf{H}(\chi,\psi):=\sup_{u_t}\{\mathrm{Re}[i\qave{\chi_t}
{\hat{H}(u_t)}{\psi_t}]-\mathsf{C}(t,u_t,\psi_t)\}$
\cite{Belavkin2009,Gough2005}, \andid{which is not to be confused with the system 
Hamiltonian $\hat H(u_t)$ that we control through $u_t$}. From the Hamilton-Jacobi-Bellman equation 
it is possible to retrieve the equations of motion for the state and the
costate, the same as in the variational approach, which look, given the
aforementioned Pontryagin Hamiltonian, formally as the Hamilton equations 
for the phase space variables $(q,p)$ in analytical mechanics. \andid{Beside this, the solution 
of the Hamilton-Jacobi-Bellman equation, defined through the solution of the latter 
(the Hamilton boundary-value problem), is given by \cite{Gough2005}: 

\begin{equation}
\mathsf{S}(t_0,\psi_0)=\mathsf{G}(\psi_T)+\int_{t_0}^T\mathrm{d}t
[
\braket{\dot\psi_t}{\chi_t} - \mathsf{H}(\chi_t,\psi_t)
],
\label{eq:Saction}
\end{equation}
which is very similar to the action in classical mechanics. We note, 
however, that the variational approach, based on the so-called 
Pontryagin maximum principle \cite{Kirk2004}, yields necessary and sufficient 
conditions for local minima, whereas dynamical programming produces results 
that are globally optimal.}

Motivated by this analogy between OC theory and analytical mechanics 
we can view the cost functional as an ``action functional''. Indeed, in analogy 
to the Hamilton's principle
where the actual evolution of a classical system is an extremum of the action
functional, which produces the well-known Euler-Lagrange equations of motion, an
extremum of $\mathsf{J}$ produces the OC equations for the state and the
costate. 

\andid{Now, let us describe how our method works. To begin with, we} 
fix the desired cost,  $\mathcal{J}$, that the system (at least) has
to attain. Since the advantage of optimizing quantum dynamics is
connected with the possibility of reaching $\ket{\psi_{\mathrm g}}$ 
by exploiting the interference of several paths in the space of
control parameters $\mathcal{U}$, \andid{we define a 
path integral in such a space.} To this aim, we 
introduce the weight $K=\andid{\int\mathfrak{D}[u_t]}e^{\frac{i}{\alpha_{\mathrm{t}}}
\mathsf{J}[u_t,\psi_t]}$ \andid{with $\mathfrak{D}[u_t]$ being some suitable measure} on the space 
$\mathcal{U}$. The form of this weight resembles the
Feynman propagator, which is motivated by the previous 
discussed analogy between analytical mechanics and OC theory\andid{, 
and by the expression (\ref{eq:Saction}) for the optimal cost.} The choice of the exponential, however, 
does not emerge from fundamental physical requirements. \refa{We think that any well-behaved function 
peaked around the optimal control pulse will allow to estimate the robustness of an optimal control problem. This conjecture is 
based on the observation, see the discussion in the following, that according to our analysis the curvature of the 
cost functional around the optimal solution is the relevant quantity to analyze.} 

The numerical factor $\alpha_{\mathrm{t}}$ \an{given in the weight $K$}, which we shall refer to 
``infidelity tolerance'', has the following property: when 
$\mathcal{J}$ approaches \andid{(from above)} the minimum of $\mathsf{J}$, 
then $\alpha_{\mathrm{t}}\rightarrow 0$. Hence, $\alpha_{\mathrm{t}}$ is the analogous of $\hbar$ in 
Feynman path integral. Indeed, when $\alpha_{\mathrm{t}}\rightarrow 0$ 
we retrieve the OC equations for the state and the costate 
we mentioned before. 
\andid{Besides this, we assume that $\alpha_{\mathrm{t}}$ exists and it is unique 
for a given OC problem, regardless of the form of the distortion 
$\delta u_t = u_t - u_t^{\mathsf{o}}$. The uniqueness is first of all a necessary condition 
or else our method would be ineffective. There is, however, a more deep reason why we 
make such an assumption. This is more easily understood in the case where $\mathsf{C}\equiv 0$ in Eq. (\ref{eq:Jdef}). 
In such a scenario the cost functional relies solely upon the state at the final time $T$, 
that is, the cost functional will depend on some time integral of $u_t$ over $[t_0,T]$ (and 
eventually on other time integrals for its time derivatives over $[t_0,T]$ 
depending on the particular control problem one is interested in). 
Thus, minimizing the cost functional means to find the right conditions for those integral 
functionals which depend only on time $T$. Hence,  the form of $u_t$ in the interval $[t_0,T)$ 
does not matter, if those functionals of the control pulse and its time derivatives fulfill the right 
conditions at the final time $T$. In this respect the infidelity tolerance has to be 
independent from the distortion we utilize, that is, it is unique. This is also what is shown by the analysis carried 
out on the examples we will discuss later in the paper.}

 \an{Even though it relies on the particular control problem we have at hand, from the above 
 outlined discussion, at least in the most relevant cases for QIP where $\mathsf{C}\equiv 0$ 
 in Eq. (\ref{eq:Jdef}), the set of EICS, $\mathcal{A}$, 
is dense, since what matters is the fulfillment of the right conditions at the final 
time $T$. For instance, in the first example we are going to consider in the next section, that is, the optimal transport 
of a particle confined in a moving harmonic trap, Ref.~\cite{Murphy2009} has showed that if 
$u_t^{\mathsf{o}}$ is the optimal solution then $\tilde{u}_t^{\mathsf{o}}=u_t^{\mathsf{o}}
+\alpha\dot{u}_t^{\mathsf{o}}$ $\forall\alpha\in\mathbb{R}$ is optimal as well (here $\mathcal{J}=0$, see also Sec.~\ref{sec:apply}). 
Thus, there is a group of optimal solutions, parametrized by the continuous variable 
$\alpha$, which is topologically dense. Similarly, this will occur for $\mathcal{J}\ne 0$, whose set of control pulses 
forms another (dense) subset $\mathfrak{U}$ in $\mathcal{U}$. Each of these possible control signals will have a 
precise spectrum that has to be within the bandwidth of $\mathfrak{U}$, namely the largest bandwidth of the elements of $\mathfrak{U}$. 
Now, if the EICS needs to have a specific bandwidth, then one has to select from $\mathfrak{U}$ 
the $u_t$ that have the right bandwidth, namely restrict (continuously) the bandwidth of $\mathfrak{U}$ 
such that the right subset of $\mathfrak{U}$ becomes $\mathcal{A}$, even though such a procedure might 
produce an empty set.
}

To be specific let us set, for the sake of simplicity, $t_0=0$ and consider only one
control pulse \andid{$u_t:[0,T]\rightarrow\mathbb{R}$, that is, only one control pulse 
is applied to the system we aim to steer during the 
time interval $[0,T]$}. We also assume that the state $\ket{\psi_t}$ obeys the 
Schr\"odinger equation with the time-dependent Hamiltonian operator $\hat{H}(u_t)$ 
and initial condition $\ket{\psi_0}\equiv\ket{\psi(0)}$. Contrarily to
(\ref{eq:Jdef}), where the state $\psi$ is an independent variable,
hereafter we render explicit the dependence of $\psi$ on the control
parameter $u_t$. Thus, we introduce the reduced cost functional
$\mathsf{J}^{\prime}[u]:=\mathsf{J}[u,\psi(u)]$. Then by performing the
Taylor expansion of $\mathsf{J}^{\prime}[u]$ around $u_t^{\mathsf{o}}$ 
to second order in $\delta u_t$ we obtain 

\andid{
\begin{eqnarray}
\mathsf{J}^{\prime}[\mathbf{u}] \simeq \mathsf{J}^{\prime}[\mathbf{u}^{\mathsf{o}}]  
+ \frac{1}{2} \delta\mathbf{u}\,\mathbf{H}\,\delta\mathbf{u}^{\mathsf{T}}
+ \mathcal{O}(\parallel\delta \mathbf{u}\parallel^3),
\label{eq:J2nd}
\end{eqnarray}
where $\parallel\cdot\parallel$ is some norm in $\mathcal{U}$, 
$\mathsf{J}^{\prime}[\mathbf{u}^{\mathsf{o}}]$ is the minimum, which, 
without loss of generality, we will set to zero, $\delta\mathbf{u}^{\mathsf{T}}$ 
is the transposed of the vector $\delta\mathbf{u}=(\delta u_2,\dots,\delta u_{N-1})$, 
and the time interval $[0,T]$ is divided in $N-1$ equal parts $\Delta t = T/(N-1)$.}
The Hessian $\mathbf{H}$ is a
real, symmetric, and positive defined matrix, and therefore it 
can be diagonalized. Additionally, we assume the boundary
conditions $\delta u_0 = \delta u_T = 0$, that is, the values $u_0$ and $u_T$ are fixed.  
\andid{We underscore that henceforth we shall work with the discretized system time evolution since in most 
cases the optimization of a given control problem is performed numerically, and 
therefore it is discretized, but most importantly, because any experiment is 
performed with a finite number of control time steps.}

\andid{Given that, let us  define a suitable norm in $\mathcal{U}$ for $\delta u_t$ by means of $K$, 
in order to obtain a quantitative appraisal of the tolerated error by a given OC scheme.} 
A suitable choice for such a norm is given by: $\ave{\mathsf{J}^{\prime}} = \vert\int
\mathfrak{D}[u_t]\mathsf{J}^{\prime}[u_t] \exp(\frac{i}{\alpha_{\mathrm{t}}}\mathsf{J}^{\prime}[u_t])\vert$ 
with $\int\mathfrak{D}[u_t]=\lim_{N\rightarrow\infty}\int\mathrm{d}u_{N-1}\dots\int\mathrm{d}u_{2}$. 
\andid{Because of the second order approximation in \an{Eq.}~(\ref{eq:J2nd}), 
the integrals appearing in $\ave{\mathsf{J}^{\prime}}$ over the control pulse at different times can 
be replaced (by making a change of integration variables) with $\delta u_k$ $\forall k=2,\dots,N-1$.} 
This norm is indeed an ``average'' in $\mathcal{U}$ of the reduced cost functional itself 
with the exponential function being a ``probability distribution''. A similar approach has 
been introduced by Rabitz \cite{Rabitz2002} where the terminal functional $\mathsf{G}$ 
in Eq. (\ref{eq:Jdef}) is averaged over a distribution function $P(\delta u_t)$. Within our 
method $P(\delta u_t)$ can be identified with the exponential function in $\ave{\mathsf{J}^{\prime}}$. 
We remark that we are not interested in some specific noise model, 
but rather to identify the portion $\mathfrak{U}\subset\mathcal{U}$ which enable us to satisfy 
$\mathsf{J}^{\prime}\le\mathcal{J}$. Basically, $\mathfrak{U}$ is determined by 
 $\ave{\mathsf{J}^{\prime}}$, which only relies on $u_t^{\mathsf{o}}$ and $\cal J$, and any 
 kind of noise has to yield pulses such that $u_t\in\mathfrak{U}$. Hence, our approach is applicable to any OC problem and cost functional 
 (\ref{eq:Jdef}). Furthermore, we note that in the next we shall consider only real-valued 
 $u_t$ (e.g., an electric current \cite{Treutlein2006}), but we underscore that the path integral can be easily 
 generalized to complex-valued $u_t$ like the amplitude and the phase of a laser field. 
 
\andid{Close to the optimal solution we can approximate
$\mathsf{J}^{\prime}$ with its second order expansion, and therefore the norm becomes}

\begin{eqnarray}
\ave{\mathsf{J}^{\prime}} \simeq 
\sqrt{\frac{\pi}{2}\alpha_{\mathrm{t}}^{3}}
\sum_{k=1}^{M}\frac{\mathcal{N}_k}{\sqrt{\lambda_k}},
\label{eq:normaF}
\end{eqnarray} 
where $\mathcal{N} _k^{-1}= \vert\prod_{j\ne k}\int\mathrm{d}\xi_je^{i\lambda_j\xi_j^2/(2\alpha_{\mathrm{t}})}\vert$ 
are normalization factors, and \an{$\lambda_k$ are the non-zero eigenvalues of $\mathbf{H}$ with $M\le N-2$.}
The above formula makes good sense, because when $\alpha_{\mathrm{t}}\rightarrow 0$
also $\ave{\mathsf{J}^{\prime}}\rightarrow 0$. 

In order to apply the above outlined formalism to some concrete example we shall
consider hereafter $\mathsf{J}^{\prime}[u_t] = 1 - \mathsf{F}(\psi_T)$, with 
$\mathsf{F}(\psi_T)=\vert\braket{\psi_{\mathrm{g}}}{\psi_T}\vert^2$. 
We note that the state $\ket{\psi_T}$ is
implicitly depending on the whole history of $u_t$ $\forall t\in [0,T)$. 
Assuming that $\mathsf{F}(u_t,\psi_t(u_t))$ is a differentiable functional of its
arguments we have 

\begin{eqnarray}
\mathsf{J}^{\prime}[u_t] =1 - \mathsf{F}(\psi_0) 
+2\int_{0}^T\mathrm{d}t \,\mathrm{Im}[
\braket{\psi_{\mathrm{g}}}{\psi_t}
\qave{\psi_t}{\hat H(u_t)}{\psi_{\mathrm{g}}}].
\label{eq:JdefOver}
\end{eqnarray}
\andid{Here we used the fact that $\mathsf{F}(\psi_T)= \mathsf{F}(\psi_0) 
+ \int_{0}^T\mathrm{d}t \frac{\mathrm{d}\mathsf{F}}{\mathrm{d}t}(\psi_t)$.} 
The most difficult part of \andid{our method} is the computation of the
  Hessian matrix $\mathbf{H}$. 
  To this aim we have
  two possibilities at our disposal: either we estimate $\mathbf{H}$
  by means of the Broyden-Fletcher- Goldfarb-Shanno \an{(BFGS)} formula 
\cite{Mordecai2003} or we compute the
  first and second derivatives of the state
$\ket{\psi_t}$ with respect to the control $u_t$. \an{Here} we choose the latter
approach, where we have to solve the following equations

\begin{eqnarray}
\partial_t\ket{\delta^2\psi} + i \hat H(u_t^{\mathsf{o}}) \ket{\delta^2\psi} &=& 
-2 i \delta\hat H(u_t^{\mathsf{o}}) \ket{\delta\psi} - i \delta^2\hat H(u_t^{\mathsf{o}})  
\ket{\psi_t^{\mathsf{o}}},\nonumber\\
\partial_t\ket{\delta\psi} + i \hat H(u_t^{\mathsf{o}}) \ket{\delta\psi} &=& 
-i \delta \hat H(u_t^{\mathsf{o}}) \ket{\psi_t^{\mathsf{o}}}, 
\label{eq:GateauxEqs}
\end{eqnarray}
which apply to any quantum closed system. Here $\ket{\psi_t^{\mathsf{o}}}$ is
the solution of the Schr\"odinger equation for $\hat
H(u_t^{\mathsf{o}})$, and $\delta\hat H(u_t^{\mathsf{o}})$, $\delta^2\hat
H(u_t^{\mathsf{o}})$ are the Gateaux derivatives of the \andid{system} Hamiltonian 
(defined as: $\delta \hat H\equiv\frac{\mathrm{d}\hat H[u_t+\alpha\delta u_t]}{\mathrm{d}\alpha}\vert_{\alpha =
  0}$). \andid{These derivatives are always analytically computable, since the dependence of the system Hamiltonian $\hat H(u_t)$ 
  on the control pulse $u_t$ is always known, whereas the analytical dependence of both $\ket{\delta^2\psi}$ 
  and $\ket{\delta\psi}$ on $u_t$ is known only in few cases (e.g., the driven and parametric harmonic oscillator \cite{Khandekar1979}). 
  Because of the latter, we need to solve the equations (\ref{eq:GateauxEqs}) which allow us to determine the matrix $\bf{H}$.
  \an{However, depending on the particular control problem, the BFGS method might be more efficient.} 
  Beside this, we note that the case of a linear quadratic regulator (or even Gaussian) control, that is, a system for which the state 
  equation is linear and the performance criterion to be minimized is a quadratic form of the state and eventually also 
  of the control pulse \cite{Kirk2004}, is not contemplated in our scheme. Indeed, the system state $\ket{\psi_t}$ is always 
  dependent on the control pulse $u_t$ through the Schr\"odinger equation 
  of motion, even in the simple scenario where the system Hamiltonian $\hat H(u_t)$ depends linearly on $u_t$. Indeed, we are 
  interested in the Hessian of the reduced cost functional $\mathsf{J}^{\prime}[u]=\mathsf{J}[u,\psi(u)]$, whose dependence on $u_t$ 
  might be not trivial through $\ket{\psi(u_t)}$. Hence, the cases in which the Hessian relies only upon 
  the state do not concern $\mathsf{J}^{\prime}[u]$.} 

We \andid{also} remark that the above outlined formalism concerns state vectors in Hilbert spaces. 
If we would be interested in the optimization of unitary transformations 
$\hat U(t) = \hat U(u_t)$, then we should perform the Taylor expansion 
of $\mathsf{F}(\hat U(T)) = \frac{1}{d}\mathrm{tr}\{\hat{U}^{\dagger}_{\mathrm{g}}\hat U(T)\}$, 
where $\hat{U}_{\mathrm{g}}$ is the ideal unitary we wish to accomplish, and 
$d$ is the dimension of the Hilbert space. Thus, the previuos analysis can be easily 
generalized to unitary operations.

\section{Applications of the method}
\label{sec:apply}

Let us first apply our method to an exactly solvable model. We consider the 
transport of a particle in a movable one-dimensional harmonic trap potential, 
for which the OC pulse and the functional dependence of the state on the controller 
are analytically known \cite{Murphy2009}.  Such control problem is of relevance for 
the realization of a quantum ion processor. Indeed, optimistic estimates show that transport 
processes may account for 95\% of the operation time of a quantum computation 
\cite{Huber2008}. Beside this, the harmonicity of the ion confinement in segmented 
Paul traps has been proven both numerically \cite{Singer2010} and experimentally 
\cite{Huber2010}.

The \andid{system} Hamiltonian is given by $\hat H(u_t) = \frac{1}{2}\{\hat p^2 +(\hat x - u_t)^2\}$ 
with $[\hat x,\hat p]=i$ (we use harmonic oscillator units). We focus our attention on the 
ground state, that is, when the particle is initially prepared in the lowest vibrational state 
of the trap. The aim is to transport such a state over the distance $\Delta x$ in a time $T$ 
such that $\psi_T(x) = e^{i\varphi}\phi_0(x-\Delta x)\equiv\psi_{\mathrm{g}}(x)$, where
$\varphi$ is an unimportant phase factor and $\phi_0(x)$ is the Gaussian harmonic
oscillator ground state wavefunction. Thus, our goal is to find a prescription such that 
when the OC pulse is perturbed the system has to reach at least the - a priori fixed - value 
of fidelity $\mathcal{F}\in[0,1]$. 

\begin{figure}[t]
\begin{center}
\includegraphics{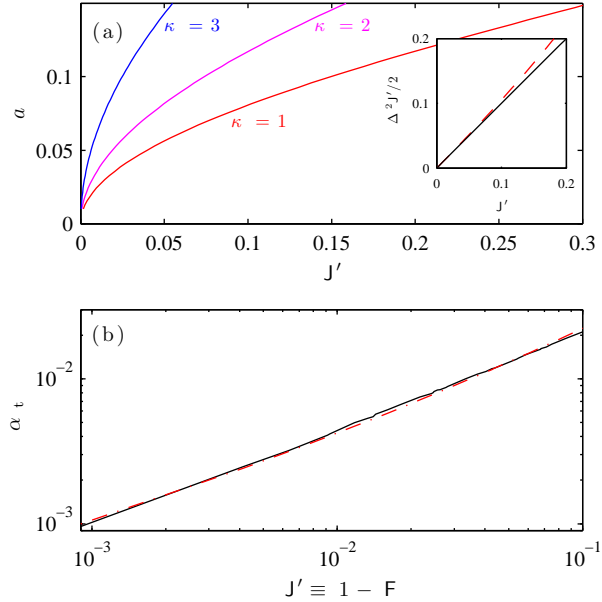}
\end{center}
\vspace{-0.7cm}
\caption{(Color online) (a) Distortion amplitude $a$ (see
  text) vs. infidelity. Inset: second order Taylor expansion of
  $\mathsf{J}^{\prime}$ \andid{($2\Delta^2\mathsf{J}^{\prime}[u_t^{\mathsf{o}},\delta
u_t,\delta u_t]\equiv\delta\mathbf{u}\,\mathbf{H}\,\delta\mathbf{u}^{\mathsf{T}}$)} 
vs. the reduced cost functional itself (red,
  $\kappa = 1$) for different values of the distortion amplitude $a$,
  whereas the black line is a guide to the eye that helps to see when
  the second order approximation holds. (b) Infidelity tolerance
  parameter $\alpha_{\mathrm{t}}$: the solid (black) line represents $[2
\lambda_{\emptyset}(\Delta^2\mathsf{J}^{\prime}[u_t^{\mathsf{o}},\delta
u_t,\delta u_t])^2/\pi]^{\frac{1}{3}}$, whereas the red (dashdot) line is a fit.}
\label{fig:Inf-vs-J2nd}
\end{figure}

In Ref. \cite{Murphy2009} the analytic solution for the time evolved state is provided. This enables
us to compute analytically the Gateaux derivatives of the state of the
system without the need of solving (\ref{eq:GateauxEqs}).
In Fig. \ref{fig:Inf-vs-J2nd}(a) we show results for a distortion 
given by:  $\delta u_t = a \sin(\kappa 2\pi t/T)
\dot{u}_t^{\mathsf{o}}$, with $u_t^{\mathsf{o}}$ being the OC pulse of 
Ref. \cite{Murphy2009} [see Eq. (5) therein]. Such simple
  distortion modulates the OC pulse at the rate $\kappa$ and 
gives a direct quantitative measure of the distortion
  degree applied to $u_t^{\mathsf{o}}$: the larger $a$ is, 
the larger the infidelity. Thus, Fig. \ref{fig:Inf-vs-J2nd}(a) shows which is the largest
admissible value of $a$ for a fixed value of the (reduced) cost
functional, whereas in the inset we show the deviation from the linear behaviour
of \andid{$2\Delta^2\mathsf{J}^{\prime}\equiv \delta\mathbf{u}\,\mathbf{H}\,\delta\mathbf{u}^{\mathsf{T}}$} for $\kappa = 1$. 

Now we write the cost functional as $2\mathsf{J}^{\prime}[u_t]
\simeq \delta\mathbf{u}\,\mathbf{H}\,\delta\mathbf{u}^{\mathsf{T}}$. 
In this specific example it turns out, numerically, that the matrix elements of 
$\bf{H}$ are of the form $H_{nk} = h + \delta H_{nk}$, 
with $h,\delta H_{nk}\in\mathbb{R}$ such that $\delta H_{nk}/h\ll 1$. This means 
that the matrix elements are almost equal to each other. The eigenvalue problem 
for such a matrix can be well approximated by $\lambda^{N-3}[\lambda - (N - 2) h] = 0$, that 
is, only an eigenvalue is non-zero. Thus, by defining $\bar{h}=\sum_{nk}H_{nk}/(N - 2)^2$ 
we get for the non-vanishing eigenvalue the simple expression 
$\lambda_{\emptyset}=(N - 2) \bar{h}$. Given these remarks, the norm
(\ref{eq:normaF}) reduces to $\ave{\mathsf{J}^{\prime}} \simeq
\mathcal{N}_{\emptyset}\sqrt{\pi\alpha_{\mathrm{t}}^3/(2\lambda_{\emptyset})}$.  

Finally, in order to determine $\alpha_{\mathrm{t}}$ we proceed in the following
way: since the norm is the ``average cost functional'' and $\alpha_{\mathrm{t}}$  
has to be unique, we simply use the inverted formula $\alpha_{\mathrm{t}} = [2
\lambda_{\emptyset}\andid{\ave{\mathsf{J}^{\prime}}}^2/(\pi\mathcal{N}_{\emptyset}^2)]^{\frac{1}{3}}$ 
for a given choice of $\delta u_t$\andid{, and perform the substitution 
$\ave{\mathsf{J}^{\prime}}\rightarrow\delta\mathbf{u}\,\mathbf{H}\,\delta\mathbf{u}^{\mathsf{T}}/2 
\simeq 1 - \mathsf{F}(\psi_T)$. 
\an{(For several non-zero eigenvalues of $\bf{H}$ we would have had $\alpha_{\mathrm{t}} = [2
\ave{\mathsf{J}^{\prime}}^2(\sum_{k=1}^M\sqrt{\pi}\mathcal{N}_k/\sqrt{\lambda_k})^{-2}]^{\frac{1}{3}}$.)}
Then, we choose some distortion $\delta u_t$ and by varying the 
strength of such a distortion we collect the values of $\alpha_{\mathrm{t}}$ versus the numerically 
exact overlap infidelities (i.e., without second order approximation), which is basically identified 
with $\mathcal{J}\equiv 1-\mathcal{F}$, the fixed error threshold}. 
We tested, however, that different kinds of distortions \an{(e.g., Fourier-like model)} produce 
practically the same curve as the one shown in Fig. \ref{fig:Inf-vs-J2nd}(b). 
This is easily understood, since what is relevant for the overlap infidelity 
terminal functional is the final state $\ket{\psi_T}$. \an{Hence, our method is general and 
it applies to any kind of distortion model. The reason for choosing the 
single frequency noise for the results displayed in 
Fig. \ref{fig:Inf-vs-J2nd}(a), was only to show the impact of the distortion on the cost 
functional in a simple and analytical way. 
}

Secondly, we perform a fit of the obtained curve for $\alpha_{\mathrm{t}}$ as a function of the overlap 
infidelity. For the present example, showed in Fig. \ref{fig:Inf-vs-J2nd}(b), we 
found that the following function well represents the data: 
$\alpha_{\mathrm{t}} = a (1-\mathcal{F}) + b \sqrt{1-\mathcal{F}}$, with $a=0.130$ and
$b=0.029$. Given that, all distortions of the optimal control pulse 
that satisfy the inequality  $\delta\mathbf{u}\,\mathbf{H}\,\delta\mathbf{u}^{\mathsf{T}}\le 
\sqrt{\pi \alpha_{\mathrm{t}}^3\mathcal{N}_{\emptyset}^2/(2\lambda_{\emptyset})} =: \ell(\mathcal{F})$
for a fixed (a priori) value of desired fidelity $\mathcal{F}$, will yield an
overlap fidelity $\mathsf{F}(\psi_T)\ge\mathcal{F}$. From an
  operational point of view, this means that if we randomly choose a
  distortion $\delta u_t$ such that $\mathcal{I}=[\delta\mathbf{u}\,\mathbf{H}\,\delta\mathbf{u}^{\mathsf{T}}
\sqrt{2\lambda_{\emptyset}/(\mathcal{N}_{\emptyset}^2\pi)}]^{2/3}/\alpha_{\mathrm{t}}\le 1-\mathcal{F}$, 
than we will certainly attain a state $\ket{\psi_T}$ whose overlap fidelity 
is greater than $\mathcal{F}$ (see also Fig. \ref{fig:sketch}). This permits 
us to find control pulses that \andid{might be} more appropriate for an experimental 
implementation. This conclusion follows from the procedure we established 
for the determination of the infidelity tolerance $\alpha_{\mathrm{t}}$. 
We underscore, however, that the threshold $\ell(\mathcal{F})$ depends only
on $u_t^{\mathsf{o}}$,  because $\mathbf{H}$ relies only on
$u_t^{\mathsf{o}}$, and that it is the result of an average in
  $\mathcal{U}$ through the path integral we defined. This provides a
  ``universal'' character to $\ell(\mathcal{F})$ for the given control problem. 

As a second example we consider the Landau-Zener model which 
has been proven useful to describe the tunneling of Bose-Einstein 
condensates in accelerated optical lattices \cite{Zenesini2009} and 
the dynamics of a quench-induced phase transition in the quantum 
Ising model \cite{Zurek2005}.  The model is described by the following 
\andid{system} Hamiltonian: $\hat H(u_t) = u_t\hat\sigma_z +
\Omega\hat\sigma_x$, where $\hat\sigma_z,\hat\sigma_x$ are Pauli
matrices. The goal is to
bring the system from the ground state $\ket{\psi_0^{\mathrm{g}}}$ of the 
Hamiltonian $\hat H(u_0)$ to the ground state $\ket{\psi_T^{\mathrm{g}}}$ 
of the Hamiltonian $\hat H(u_T)$ through the avoided level crossing. 
As objective functional we consider  $\mathsf{J}^{\prime}[u_t] = 1 - 
\mathrm{Re}[\braket{\psi_T^{\mathsf{g}}}{\psi_T}]$, that is, we also 
control the phase of the state. \andid{Even though the analytical 
dependence of the system Hamiltonian $\hat H(u_t)$ is known, 
and therefore the Gateaux derivative ($\delta\hat{H} = \hat\sigma_z \delta u_t$),} 
for such a control problem both the 
$u_t^{\mathsf{o}}$ and the dependence of $\ket{\psi_t}$ on $u_t$ 
are not analytically known. Concerning the latter, this implies that is not possible to compute
analytically the Gateaux derivatives $\ket{\delta\psi}$ and $\ket{\delta^2\psi}$, and therefore the Hessian $\bf{H}$. 
Hence, we need to solve (\ref{eq:GateauxEqs}). Given that, we seek an $u_t^{\mathsf{o}}$ by using the Krotov iterative method 
\cite{Krotov1996,Montangero2007}. We then proceed on, as in the former example, by determining 
the infidelity tolerance $\alpha_{\mathrm{t}}$, for which we get a similar fit  $\alpha_{\mathrm{t}} = 
a (1-\mathcal{F}) + b (1-\mathcal{F})^{c}$ with $a = 0.016$, $b = 0.047$, and $c = 0.650$. The criterion to be satisfied 
is then again given by:
$\delta\mathbf{u}\,\mathbf{H}\,\delta\mathbf{u}^{\mathsf{T}}\le\ell(\mathcal{F})$,
but with a different numerical value for $\ell(\mathcal{F})$.

\begin{figure}[t]
\begin{center}
\includegraphics[scale=1.00]{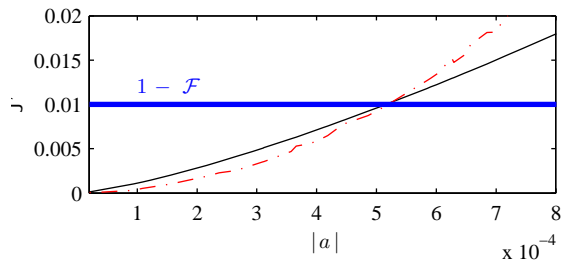}
\end{center}
\vspace{-0.7cm}
\caption{(Color online) Cost functional for the Landau-Zener model for 50
  random realizations of the distortion $\delta u(t)=a \sin(\kappa 2\pi t/T)
\dot{u}^{\mathsf{o}}(t)$: the
  red (dashdot) line is the exact infidelity, the black (solid) line correspondes to
  the second order approximation $\mathcal{I}$ (see text), and
  the blue (thick) line is the infidelity threshold $1 - \mathcal{F}= 0.01$. }
\label{fig:CheckLZ}
\end{figure}
In Fig.~\ref{fig:CheckLZ} it is showed the exact result of the
infidelity (red) and the second order approximation $\mathcal{I}$ 
(black) for 50 random realizations of $\delta u_t$ for a distortion
similar to the one of Fig. \ref{fig:Inf-vs-J2nd}(a). Instead, the
desired upper limit of infidelity $1-\mathcal{F} = 0.01$, that the
system has at least to attain, is represented by the horizontal blue line. 
Analogously to the former example, only the realizations of $\delta u_t$ that fulfil
$\delta\mathbf{u}\,\mathbf{H}\,\delta\mathbf{u}^{\mathsf{T}}\le\ell(0.99)$
have an infidelity below 1\%. \an{
As for the previous example, the single frequency model has been adopted for 
convenience in order to illustrate the effectiveness of our approach as depicted in Fig.~\ref{fig:CheckLZ}. 
Such a choice, however, does not invalidate our methodology, which we have tested 
for a Fourier-like distortion model (i.e. a sum over a finite number of harmonics 
with different amplitudes), similarly to the above outlined determination of the infidelity tolerance.
}

In general, in order to identify robust control pulses amenable to an experimental implementation 
one should proceed as follows: 1) determine the optimal control $u_t^{\mathsf{o}}$ 
with some numerical optimization algorithm; 2) compute the eigenvalues of the 
Hessian matrix $\mathbf{H}$; 3) determine the infidelity tolerance $\alpha_{\mathrm{t}}$ 
as described in the \andid{first example}; 4) randomly generate distortions $\delta\mathbf{u}$ 
such that the condition \an{$\mathcal{I}=[\sqrt{2/(\pi\alpha_{\mathrm{t}}^3)}(\sum_{k=1}^{M}\mathcal{N}_k/\sqrt{\lambda_k})^{-1}
\,\delta\mathbf{u}\,\mathbf{H}\,\delta\mathbf{u}^{\mathsf{T}}]^{2/3}\le \mathcal{J}$} 
is fulfilled for a fixed (a priori) value of $\mathcal{J}$; 5) define the new control pulse 
as $\mathbf{u} = \mathbf{u}^{\mathsf{o}} + \delta\mathbf{u}$; 6) choose the most 
suitable control pulse to be employed in the laboratory from the ensemble $\cal{E}=\{\mathbf{u}\}_{\cal 
\le \mathcal{J}}$. 
\andid{We note, however, that such a ``recipe" does not guarantee that we are able to obtain 
with certainty a control pulse more amenable for use in an experiment, but at least 
it helps to design new ones that are still able to yield a value of the cost functional below the fixed 
threshold $\mathcal{J}$. In other words, the set $\cal{E}\cap\cal{A}$ might be empty, and the elements of $\cal{E}$ are 
not necessarily close to the optimal $\mathbf{u}^{\mathsf{o}}$ (e.g., see Ref. \cite{Murphy2009}).}
In this respect, such a procedure \andid{might} help the quest of both robust and experimentally 
feasible pulses for controlling different quantum phenomena. 

\section{Conclusions}

In conclusion we have presented a method to assess the \andid{error} of
solutions to OC problems. The method might be a helpful tool for experimentalists 
in order to design experiments robust against source of noise. For instance, to estimate
how much the schemes for realizing quantum gates are robust against imperfections 
of the optimal pulse shape. 
\an{
Compared to other methods, such as the one of Ref. \cite{Shuang2004}, our technique does not need 
the simulation of a large ensemble of samples in order to minimize on average both the mean 
and the variance of the objective functional. Instead, once the Hessian matrix $\bf{H}$ and the 
infidelity tolerance $\alpha_{\mathrm{t}}$ are known, which only rely on the optimal 
(known) control pulses, one has simply to randomly generate the distortion $\delta\mathbf{u}$ and 
perform the matrix multiplication $\delta\mathbf{u}\,\mathbf{H}\,\delta\mathbf{u}^{\mathsf{T}}$, which 
is a less demanding computational task than the application of a genetic algorithm, as the one 
proposed in Ref.~\cite{Shuang2004}. Besides this, our method can be applied to both known 
error models and to evaluate unknown errors such as random telegraphic noise. 
For the future, we plan to extend our formalism to dissipative quantum systems
  (e.g., systems governed by the Born-Markov master equation).}
  
\section*{Acknowledgments}

We acknowledge financial support by  the IP-SOLID and the National 
Research Foundation and Ministry of Education Singapore (R.F.),
IP-AQUTE and PICC (T.C.), SFB/TRR21 (A.N.,T.C.), the Marie Curie program
of the European Commission (Proposal No. 236073,
OPTIQUOS) within the 7th European Community Framework
Programme and the Forschungsbonus of the University of Ulm
and of the UUG (A.N.). A. N. acknowledges  
conversations with K. Urban, J. T. Stockburger, and I. Kuprov.


%

\end{document}